# On the properties of the phenomenon like Vavilov-Cherenkov one with emission of surface plasmons

V.S.Zuev
The P.N.Lebedev Physical Institute of RAS
119991 Moscow

G.Ya.Zueva
The A.M.Prokhorov General Physics Institute of RAS
119991 Moscow
vizuev@sci.lebedev.ru

The phenomenon is described that is an analogue of the Vavilov-Cherenkov phenomenon and that could be observable in a space with the boundary separated a well reflecting light metal and a dielectric. An electron that is moving at a constant velocity in proximity to the such boundary creates surface plasmon-polaritons with a phase velocity that is below the electron velocity. The linear momentum of a photon in the created plasmon-polariton is calculated. This momentum could be as high as a doubled initial electron momentum. In the other case the electron that has created the photon in the plasmon-polariton could totally lost its velocity. It is shown that at the non-zero temperature this process has a non-zero probability.

## Особенности явления типа Вавилова-Черенкова с испусканием поверхностных плазмонов

В.С.Зуев
Физический ин-т им. П.Н.Лебедева РАН
119991 Москва

Г.Я.Зуева
Институт общей физики им. А.М.Прохорова
119991 Москва
vizuev@sci.lebedev.ru

Рассмотрено явление типа Вавилова-Черенкова, которое можно наблюдать в пространстве с границей раздела хорошо отражающий свет металл-диэлектрик. Электрон, равномерно движущийся вблизи границы раздела, порождает поверхностные плазмон-поляритоны с фазовой скоростью, которая меньше скорости электрона. Вычислен импульс фотона в возбуждаемом плазмон-поляритоне. Этот импульс может быть почти равен удвоенному начальному импульсу электрона. Электрон, породивший фотон в плазмон-поляритоне, может также практически остановится. Показано, что при не равной нулю температуре этот процесс имеет ненулевую вероятность.

# Особенности явления типа Вавилова-Черенкова
## с испусканием поверхностных плазмонов

В.С. Зуев
Физический ин-т им. П.Н. Лебедева РАН
119991 Москва

Г.Я. Зуева
Институт общей физики им. А.М. Прохорова
119991 Москва
vizuev@sci.lebedev.ru

Явление Вавилова-Черенкова – излучение света при равномерном движении быстрого электрона в среде наблюдают как в однородных средах /1/, так и в неоднородных средах, в таких, как фотонные кристаллы /2/. Условием возникновения этого излучения является наличие в пространстве (однородном или неоднородном) собственных электромагнитных волн с фазовой скоростью меньше, чем скорость возбуждающего электрона.

В пространстве с границей раздела металл-диэлектрик существуют поверхностные волны - поверхностные плазмон-поляритоны, собственные волны с малой фазовой скоростью. Для краткости будем называть их плазмонами. Отличие фазовой скорости этих волн от скорости света в вакууме может достигать многих десятков и сотен раз. Поэтому испускать излучение в виде плазмона будет электрон, сравнительно медленный по сравнению с электроном, способным излучать в однородной среде, см. /3/ и цитированную там литературу.

Энергия возбуждаемых фотонов в плазмонах, как будет видно из дальнейшего, составляет приблизительно $5\,eV$. Процесс испускания фотонов приводит к эффекту отдачи, что проявляется в уменьшении энергии электрона и в изменении его импульса. Предметом данной статьи как раз и является вычисление этих изменений.

Рассмотрение эффекта отдачи проведем так, как это сделано в /4/. Отличия будут заключаться в рассмотрении поверхностных волн.

Будем исходить из законов сохранения энергии и импульса. Будем рассматривать нерелятивистский электрон $v_e/c \ll 1$.

$$\frac{m_e v_e^2}{2} = \frac{m_e v_{e'}^2}{2} + \hbar\omega', \tag{1}$$

$$m_e \vec{v}_e = m_e \vec{v}_{e'} + \hbar\vec{k}'. \tag{2}$$

Здесь $m_e$, $\vec{v}_e$ и $\vec{v}_{e'}$ - масса электрона, его начальная и конечная скорости, $\omega'$ и $\vec{k}'$ - частота и волновой вектор испущенного фотона. Угол между первоначальной скоростью электрона $\vec{v}_e$ и волновым вектором $\vec{k}'$ испущенного фотона обозначим с помощью $\theta$, $\cos\theta = \vec{k}'\cdot\vec{v}_e / k'v_e$. Из (1), (2) получаем

$$k' = \frac{m_e v_e}{\hbar}\cos\theta\left[1 \pm \sqrt{1 - \frac{2\hbar\omega'}{m_e v_e^2 \cos^2\theta}}\right]. \tag{3}$$

Поскольку в (3) $k'$ - это модуль волнового вектора, то есть величина определенно положительная, то должно быть $\cos\theta \geq 0$. Вылет фотона может происходить в полусферу в направлении первоначального движения электрона.

Рассмотрим случай $\cos\theta = 1$. Первоначально будем считать, что $2\hbar\omega'/m_e v_e^2 \ll 1$. В зависимости от выбора знака перед корнем в (3) импульс фотона оказывается приближенно равным либо

$$\hbar k' \approx 2m_e v_e, \tag{4}$$

либо
$$\hbar k' \approx \hbar \omega'/v_e. \qquad (5)$$

В первом случае – определенно, а во втором случае – скорее всего, импульс фотона превышает и значительно импульс фотона той же частоты в вакууме.

Большой импульс означает малую скорость волны. В однородном пространстве едва ли найдутся волны необходимо малой скорости. В пространстве с границей раздела диэлектрик–хорошо отражающий металл такие волны существуют. Это плазмон–поляритоны на границе раздела металл–диэлектрик, см. /5/ и цитированную там литературу.

Дисперсионные кривые поверхностных плазмонов показаны на рис.1. Кривая 1 - для симметричного плазмона, кривая 2 – для антисимметричного плазмона, оба на пленке толщиной 10 $nm$ из металла с плазменной частотой $\omega_{pl} = 1.346 \cdot 10^{16}\, rad/s$ в вакууме, кривая 3 – для плазмона на одиночной границе раздела для того же металла и вакуума, прямая 4 проведена на уровне $\omega = \omega_{pl}/\sqrt{2}$, прямая 5 – световая линия $\omega = kc$, пунктирная прямая – функция $kv_e$. Участки кривых 1, 2 или 3, лежащие правее точек пересечения с прямой $kv_e$, дают значения волновых чисел $k'$ и частот $\omega'$ плазмонов, которые могут быть возбуждены электроном со скоростью $v_e$.

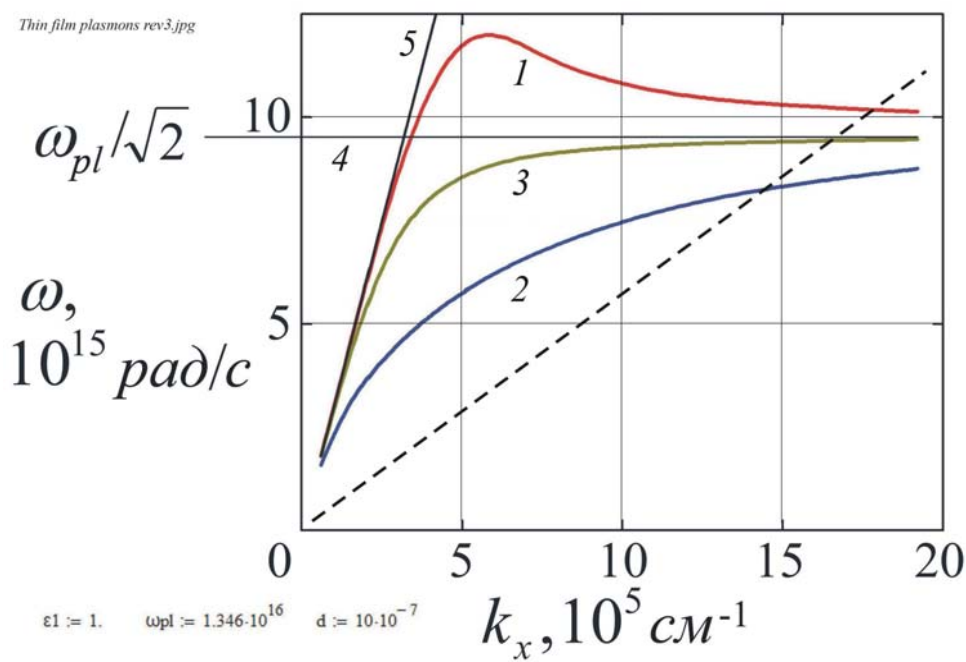

Рис.1.

При $v_e < c$ значения частот плазмонов близки к величине $\omega_{pl}/\sqrt{2}$, $\hbar\omega_{pl}/\sqrt{2} \approx 6.3\, eV$, что мы будем учитывать в дальнейшем.

Волновое число плазмона может быть сколь угодно большим (в рамках принятой модели металла – среда с отрицательным $\varepsilon$). Его частота приблизительно равна $\omega_{pl}/\sqrt{2}$.

В случае, описываемом формулой (4), импульс испущенного фотона приблизительно равен удвоенному начальному импульсу электрона. Электрон, испустив такой фотон, поворачивает назад. Его энергия меньше первоначальной на величину энергии испущенного фотона, соответственно меньше по абсолютной величине его импульс, направленный теперь строго в обратную сторону.

В случае, описываемом формулой (5), фотон по-прежнему имеет значительный импульс, хотя и не такой большой, как по формуле (4). Значит, это снова поверхностный плазмон. Электрон, испустив фотон, продолжает двигаться в прежнем направлении. Частота испущенного фотона приблизительно равна $\omega_{pl}/\sqrt{2}$.

Теперь возьмем более медленный электрон, такой, что $\frac{m_e v_e^2}{2} \approx \hbar\omega_{pl}/\sqrt{2}$, хотя и больше последнего. Это может быть электрон проводимости в металле. Мы, однако по-прежнему будем рассматривать электрон вне металла в малой близости к его поверхности. Если бы мы рассматривали электрон в металле, то потребовалось бы рассмотрение явления параметрического распада электронной волны на вторичную электронную волну и на электромагнитную волну, чего мы пока не делаем.

По-прежнему выберем $\cos\theta = 1$. При $\frac{2\hbar\omega'}{m_e v_e^2} \approx 1$ малой величиной в (3) оказывается $\sqrt{1 - \frac{2\hbar\omega'}{m_e v_e^2 \cos^2\theta}}$. Из двух возможностей в (3) следует выбрать знак минус, так как при выборе плюса импульс фотона оказывается по абсолютной величине больше начального импульса электрона.

Электрон, испустив фотон, практически останавливается. Для электрона вне металла это вполне возможно. А для электрона в металле этот процесс на первый взгляд представляется невозможным: в вырожденном газе электронов в металле нет вакантного места для дополнительного электрона с нулевой скоростью.

На самом деле это не так. При температуре $T > 0$ фермиевская ступенька в функции распределения электронов по энергии оказывается размытой. При равной нулю энергии имеется экспоненциально малая, но конечная вероятность существования вакансии. Эту вакансию и будет занимать электрон, потерявший свою энергию и породивший фотон с энергией и импульсом, равными начальной энергии и начальному импульсу исходного электрона. Величина вероятности перехода в такое вакантное состояние не будет определяться величиной вероятности существования вакансии, а будет определяться плотностью состояний для вакансии.

В пространстве с металлической нитью быстрый электрон будет возбуждать цилиндрические волны, простирающиеся по радиусу до бесконечности, и поверхностные волны. Медленный электрон будет возбуждать только поверхностные волны.

Рассмотрим численный пример. Пусть начальная энергия электрона равна $E = 50\ eV$ ($8 \cdot 10^{-11}\ erg$). Скорость такого электрона равна $v_e = 3 \cdot 10^8\ cm/s$, импульс $m_e v_e = 2.7 \cdot 10^{-19}\ g \cdot cm/s$. Энергия испускаемого фотона $\hbar\omega'$ будет приблизительно равна $\hbar\omega_{pl}/\sqrt{2} = 6.3\ eV$. По формуле (3) с плюсом перед корнем и при $\cos\theta = 1$ волновое число оказывается равным $k' = 4.9 \cdot 10^8\ cm^{-1}$, скорость фотона $v'_{ph} = \frac{\omega'}{k'} = 2.7 \cdot 10^7\ cm/s$, что, как и должно быть, меньше $v_e$, импульс фотона $\hbar k' = 5.2 \cdot 10^{-19}\ g \cdot cm/s$, то есть равен почти удвоенному начальному импульсу электрона. Электрон, испустив такой фотон, будет двигаться в обратном направлении.

Подведем итог. Электрон, равномерно движущийся вблизи поверхности раздела металла и диэлектрика, возбуждает поверхностные плазмоны. По существу, это явление аналогично явлению Вавилова-Черенкова в однородной среде. Существенное отличие заключается в том, что в процессе с плазмонами электрон может иметь заметно меньшую скорость в сравнении с электроном в однородной среде.

Среди порождаемых фотонов есть такой, импульс которого почти равен удвоенному начальному импульсу электрона. Испустив такой фотон, электрон будет двигаться почти с прежней скоростью, но в обратном направлении. В однородной среде такой процесс невозможен.

Процесс с рождением электронов обратного направления движения может быть применен в устройстве регистрации медленных электронов.

1. Л.Д.Ландау, Е.М.Лифшиц. Электродинамика сплошных сред. Москва, Наука, 1982
2. C.Luo, M.Ibanescu, S.G.Johnson, J.D.Joannopoulos. Science, v.299, pp.368-371 (2003)
3. В.С.Зуев, А.М.Леонтович, В.В.Лидский. Черенковский механизм возбуждения поверхностных волн. Оптика и спектроскопия, т.110, №3, 446-452 (2011)
4. В.Л.Гинзбург. Некоторые вопросы теории излучения при сверхсветовом движении в среде. УФН, т.69, 537-564 (1959)
5. В.С.Зуев, Г.Я.Зуева. Очень медленные поверхностные плазмоны: теория и практика. Оптика и спектроскопия, т.107, №4, 648–663 (2009)